# New insights into the mechanism of graphene oxide and radionuclide interaction through vacancy defects

*Anastasiia S. Kuzenkova[1], Anna Yu. Romanchuk[1], Alexander L. Trigub[2], Konstantin I. Maslakov[1], Alexander V. Egorov[1], Lucia Amidani[3,4], Carter Kittrell[5,6], Kristina O. Kvashnina[3,4], James M. Tour[5,6], Alexandr V. Talyzin[7] , Stepan N. Kalmykov[1, *]*

[1] - Department of Chemistry, Lomonosov Moscow State University, Leninskie Gory, Moscow 119991, Russia. stepan@radio.chem.msu.ru

[2] - National Research Centre "Kurchatov Institute", Moscow, Russia

[3] - Rossendorf Beamline at ESRF – The European Synchrotron, CS40220, 38043 Grenoble Cedex 9, France,

[4] - Helmholtz Zentrum Dresden-Rossendorf (HZDR), Institute of Resource Ecology, PO Box 510119, 01314 Dresden, Germany

[5]- Department of Chemistry, Smalley-Curl Institute and Nanocarbon Center, PO Box 1892, MS-100 Houston, Texas, 77251, USA

[6]- Department of Materials Science and Nano Engineering, Rice University, 6100 Main St, Houston, Texas, 77005, USA

[7] – Department of Physics, Umeå University, S-90187, Umeå, Sweden.

ABSTRACT The sorption of U(VI), Am(III)/Eu(III) and Cs(I) radionuclides by graphene oxides (GOs) synthesized by Hummers's, Brodie's and Tour's methods was studied through a combination of batch experiments with characterization by microscopic and spectroscopic techniques such as XPS, ATR-FTIR, HERFD-XANES, EXAFS and HRTEM. Remarkably different sorption capacity and affinity of radionuclides was found towards GOs synthesized by Hummers's and Brodie's methods reflecting different structure and oxidation state of these materials. Mechanism underlying GO – radionuclide interaction is determined using variety of experimental techniques. For the first time it is shown here that GO - radionuclides interaction takes place on the small holes or vacancy defects in the GO sheets. Mechanism of GO's interaction with radionuclides was analyzed and specific functional groups responsible for this interaction were identified. Therefore, a new strategy to produce improved materials with high capacity for radionuclides suggests the use perforated and highly defected GO with a larger proportion of carboxylic functional groups.

---

* Corresponding author. E-mail: stepan@radio.chem.msu.ru. Tel: +7-495-939-3220

## 1. INTRODUCTION

Nuclear power is one of the most important energy production methods. Presently, 32 countries operate 194 nuclear power plants with ~450 blocks. The average worldwide share of nuclear energy is c.a. 11% with much larger numbers in several countries, e.g. as high as 72% in France [1]. Nuclear energy production results in significantly decreased carbon dioxide emission to the atmosphere compared to traditionally used fossil fuels [2]. However nuclear energy faces the challenge of radioactive waste management. The processing of the radioactive waste aims at reducing their volume and converting them into forms convenient for long-term disposal [3]. To solve this problem different sorbent materials have been studied [4,5], including zeolites [6–8], cement-based materials [8,9], clays [10–12], carbon materials [13–18] etc. Special attention was given more recently to promising carbon nanomaterials and especially to graphene oxide (GO). GO is one layer of carbon with various functional oxygen-containing groups (epoxy, carboxyl, carbonyl, phenolic, hydroxyl etc.) [19,20] It is an easy to synthesize GO, it's relatively inexpensive, non-toxic and effective. As 2D nanomaterial it has a high specific surface area, e.g. experimentally determined values to be ~ 700-800 $m^2/g$ [21,22] while the theoretical values are as high as 2600 $m^2/g$ for a single graphene sheet [23]. The high sorption ability of GO towards various cations of radionuclides and heavy metals was repeatedly demonstrated previously [24–34]. Various mechanisms of sorption of radionuclides onto GO have already been proposed, however, little is known about the chemistry behind the cation-GO interaction, i.e. which oxygen-containing groups of GO bind these cations and how these groups are localized on the GO sheet.

There are several methods to produce GO which provide distinctly different materials with a strong difference in relative abundance of various functional oxygen-containing groups. The most common of them are Hummers's (HGO) and Brodie's (BGO) methods. HGO [35] involves oxidation of graphite by $KMnO_4$ in solutions of $NaNO_3$ and $H_2SO_4$. BGO is made using fuming nitric acid and $NaClO_3$ [36]. Earlier it was shown that GO synthesized by HGO and BGO methods have different relative amounts of oxygen groups and significantly different swelling properties. [37] BGO has fewer defects and more homogeneous distribution of functional groups over its surface. HGO shows a relatively high percentage of carbonyl and carboxyl groups with a significant number of holes in the flakes and a strong disruption of the graphene structure. Tour's method (TGO) is the more recent modification of the Hummer's method and is the oxidation of graphite by $KMnO_4$ in a mixture of $H_2SO_4$ and $H_3PO_4$ in a ratio of 9:1 [38]. TGO is typically more oxidized compared to HGO.

It can be anticipated that the sorption of radionuclides will be different for GO prepared by different methods reflecting the variations in relative amounts of different functional groups. However, so far the sorption of radionuclides was reported only for HGO and TGO. No systematic study of sorption properties vs. composition of GO is available in literature and a detailed mechanism of radionuclide interactions with GO remains unclear. Understanding of the mechanism could help to design new improved GO-based sorbents for radionuclide waste treatments.

The aim of this study is to reveal mechanisms of radionuclide sorption by various GOs, to determine which of the oxygen-containing groups preferentially interact with radionuclides and how the adsorbed radionuclides are distributed over the surface. In this study, radionuclides with different chemical properties were chosen. In particular, U(VI), Cs(I), Am(III) and its analogue Eu(III) were studied. Also, [137]Cs, [241,243]Am and various U isotopes are of primary interest in the field of nuclear waste management and disposal as well as remediation of nuclear legacy sites Sorption of radionuclides is compared for BGO, HGO and TGO. The sorption of various radionuclides was done with U(VI), Eu(III)/Am(III) and Cs(I). Batch sorption experiments were characterized by spectroscopic and microscopic methods, namely X-ray photoelectron spectroscopy (XPS), attenuated total reflection fourier-transform infrared spectroscopy (ATR-FTIR), high-energy resolution fluorescence detected X-Ray absorption spectroscopy (HERFD-XANES), extended X-ray absorption fine structure (EXAFS) and high resolution transmission electron microscopy (HRTEM).

## 2. EXPERIMENTAL

### 2.1 Materials

BGO was performed using 5 g of small flake graphite (Graphexel, <200 μm). It was mixed with 42.5 g of sodium chlorate, placed in an ice bath and 30 mL of fuming nitric acid was added dropwise over ~1 h time under continuous stirring. The mixture was continuously stirred for c.a 12 h. at ambient temperature and then heated to 60 °C for 8 h. After repeated washing with deionized water and 10% HCl solution, the paste was freeze-dried to get a brown-colored BGO powder.

HGO was purchased from ACS Materials (CAS No.: 7782-42-5). GO produced by TGO was donated by Zonko LLC, USA.

### 2.2 Sorption experiments

Sorption experiments were carried out in plastic vials while retention on the walls was found to be negligible. To prepare the suspension, GO powder was dissolved in water and treated by ultrasound in a cavitation mode. For sorption experiments, an aliquot of the solution containing the radionuclides ($^{241}$Am, $^{233,232}$U, $^{137}$Cs) was added to a 0.07 g/L GO suspension in a 0.01 M NaClO$_4$ (experiments with each radionuclide were performed separately). The pH value was measured using a combined glass pH electrode (InLab Expert Pro, Mettler Toledo) with an ionomer (SevenEasy pH S20-K, Mettler Toledo) and was adjusted via addition of small amounts of dilute HClO$_4$ or NaOH. After equilibration, the GO suspension was centrifuged at 40,000 g for 20 min (Allegra 64R, Beckman Coulter) to separate the solid phase from the solution. The sorption was calculated using the difference between the initial activity of the radionuclides and the activity measured in the solution after centrifugation. The activity of the radionuclides was measured using liquid-scintillation spectroscopy (TriCarb 2700TR, Canberra Packard Ind., USA and Quantulus-1220, Perkin Elmer) and universal radiometric complex ORTEC DSPec50 (16013585).

Experiments in the saturation mode were carried out with all samples. In this case a competing cations were added to the GO suspension, i.e. [Al$^{3+}$] = 1.3g/L for Am(III) experiment, [Ca] = 1.3g/L for U(VI) experiment and [Na] = 0.2 g/L for Cs(I) experiment. Concentration of all GO samples in these experiments was kept at 0.7 g/L.

### 2.3 Sample preparation for spectroscopic and microscopic characterization.

For safety, stable or long-lived isotopes were used in the preparation of samples for characterization using spectroscopic and microscopic techniques. In the case of U(VI), natural uranium was used. In experiments that simulated Am(III) behavior, its chemical analogue Eu(III) was used. A stable isotope of Cs(I) was used instead of $^{137}$Cs. The list of the prepared samples together with details of the experiments are presented in Table S1.

### 2.4 Methods of characterization

XPS spectra were acquired on an Axis Ultra DLD spectrometer (Kratos Analytical Limited, Great Britain) with a monochromatic Al Kα radiation (hv = 1486.6 eV, 150 W). The pass energy of the analyzer was 160 eV for survey spectra and 40 eV for high resolution scans. The GO samples were mounted on a holder using a double-sided adhesive tape. The Kratos charge neutralizer system was used, and the spectra were charge-corrected to give the main component of O1s peak a binding energy of 532.5 eV. Spectra were curve fitted using CasaXPS software.

ATR-FTIR spectra were obtained using a PLATINUM, a single reflection horizontal ATR accessory from Bruker Technologies, equipped with a diamond crystal. Ungrounded small species of the samples were analyzed at room temperature. For each sample, 256 scans were recorded under vacuum in the MIR region (4000-400 $cm^{-1}$) with a resolution of 4 $cm^{-1}$.

The U $L_3$-edge and Eu $L_2$-edge EXAFS spectra were collected at the Rossendorf Beamline BM20 of the European Synchrotron Radiation Facility (ESRF), Grenoble, France. The incident energy was selected using the <111> reflection from a double water-cooled Si crystal monochromator. Rejection of higher harmonics was achieved by two Rh mirrors at an angle of 2.5 mrad relative to the incident beam. The incident X-ray beam had a flux of approximately 2 x $10^{11}$ photons $s^{-1}$ on the sample position. XAFS data were recorded in fluorescence mode using a 13-element high-throughput Ge-detector. The recorded intensity was normalized to the incident photon flux. Data were collected up to k = 10 $A^{-1}$ with a typical acquisition times of 20 min per spectrum.

EXAFS data ($\chi exp(k)$) were analyzed using the IFEFFIT data analysis package [39]. Standard procedures for the pre-edge subtraction and spline background removal were used for EXAFS data reduction. The radial pair distribution functions around the U/Eu ions were obtained by the Fourier transformation (FT) of the k2-weighted EXAFS functions $\chi exp(k)$ over the ranges of photoelectron wave numbers k = 2.5–12.0 $Å^{-1}$. The structural parameters, including interatomic distances ($R_i$), coordination numbers ($N_i$) and Debye–Waller factors ($\sigma 2$), were found by the non-linear fit of theoretical spectra (eq. 1) to experimental ones:

$$(1)$$

The theoretical data were simulated using the photoelectron mean free path $\lambda(k)$, amplitude $F_i(k)$ and phase shift $\varphi_i(k)$ calculated ab initio using program FEFF6. Experimental spectra were fitted in R-space within range 1.2-4.2 Å. For the refined interatomic distances ($R_i$), the statistical error is 0.01–0.02 Å for the first coordination sphere.

XANES spectra in high-energy-resolution fluorescence detection (HERFD) mode on the U-GO samples were recorded at the Rossendorf Beamline (BM20) using an X-ray emission spectrometer [40]. The sample, analyzer crystal and photon detector (silicon drift diode) were arranged in a vertical Rowland geometry. The U HERFD spectra at the L3 edge were obtained by recording the maximum intensity of the U $L_{III}$ emission line (13616 eV) as a function of the incident energy. The emission energy was selected using the <777> reflection of one spherically bent Ge crystal analyzers (with 0.5 m bending radius) aligned at 77° Bragg angle. The intensity was normalized to the incident flux. A combined (incident convoluted with emitted) energy resolution of 3.0 eV was obtained as determined by measuring the full width at half maximum (FWHM) of the elastic peak.

The HRTEM images were obtained with an aberration-corrected JEOL 2100F system operated at 200 kV, yielding an information limit of 0.8 Å. The diffraction images and energy-dispersive X-ray spectroscopy (EDX) data were obtained in scanning transmission electron mode, with the spot size of 1 nm with the HAADF and JED 2300 (JEOL) detectors. The samples for the characterization were prepared on the TEM copper grid. For this purpose, the GO suspension was first drop cast onto the grid. After that, Eu(III) $10^{-5}$ M solution at pH 5 was drop cast atop and left for ~5 s. After that the excess solution was removed, the grid with sample was washed by water and air dried.

### 2.5 DFT geometry optimization

Local atomic geometries were investigated by using ab initio simulations based on density functional theory (DFT+U) in an unrestricted open-shell Kohn−Sham framework, as implemented in the Quantum ESPRESSO [41]. Atomic clusters have been placed into large periodic supercells and fully optimized until the forces on all nuclei were smaller than 0.001 au (Rydberg/Bohr) and until an energy difference smaller than 0.0001 Ry was calculated between two steps of the minimization algorithm. The simulations have been carried out by using the Γ-point for the k-point sampling of the Brillouin zone, norm-conserving PAW pseudopotentials generated using a Martin-Troullier scheme [42] (distributed by Davide Ceresoli https://sites.google.com/site/dceresoli/pseudopotentials) with nonlinear core correction and the PBE exchange−correlation functional. Kohn−Sham orbitals have been expanded into plane waves up to energy cutoffs of 70 Ry for the wave functions in order to achieve satisfactorily converged results.

The optimized structures have been used to simulate HERFD-XANES U $L_3$-edge spectra using the FDMNES code [43]. Spin-orbital and relativistic effects were considered and the Green's function method was used. A cluster of 6.5 Å radius around the absorber, including up to 57 atoms, was considered.

### 3. RESULTS AND DISCUSSION

### 3.1 Characterization of GO materials.

All samples of GOs were characterized by XRD, XPS and IR-spectroscopy. It is well known that the structure of GO depends on the synthesis method, the details and conditions of synthesis procedure and even on the starting graphite material [44,45] Therefore, extensive characterization of the pristine material is required to understand the main factors affecting the outcome of the sorption experiments. As established using XPS data (Table S2, Fig.S1A) TGO and HGO are more oxidized (C/O = 1.85 and C/O = 2.37, respectively) compared to the BGO (C/O = 2.59). HGO and TGO samples exhibited also a small amount of sulfur impurities. The sulfate groups is the common impurity for GO synthesis methods which came from the sulfuric acid [46,47] Assuming that all sulfur originate from sulfate groups, the impurity corrected C/O ratios for TGO and HGO are 2.09 and 2.66 respectively. Therefore, HGO and BGO samples demonstrate nearly identical oxidation degree while TGO sample is more heavily oxidized. Fitting of the C1s XPS spectra also showed different relative amount of oxygen functional groups (Table S3). The spectra were fitted with three components assigned to C-O, C=O and C-C carbons according to most common procedure [48] and showed notably smaller amount of C=O functional groups in BGO. Moreover, the different distribution of oxygen groups for the different GOs is confirmed by analysis of FTIR spectra. The TGO and HGO samples demonstrate much higher content of carboxyl groups compared to the BGO (Fig. S2). This result is in agreement with previously published data [37,49,50].

XPS also provides rough estimate for defect state of GO. Formation of C=O is possible if the carbon atom is bonded only to two carbon neighbors. Therefore, each C=O provides point defect. Considering that size of GO flakes is on micrometer scale, the number of edge atoms is negligibly small compared to the total number of carbon atoms. Therefore, most of the C=O (up to ~7% for HGO) must be placed on planar surface and originated from holes and small defects. Recent FTIR imaging experimental data confirm that C=O groups are distributed over the whole surface of GO [51]. Notably, XPS shows almost twice higher amount of C=O in HGO and TGO compared to less defect BGO in agreement with earlier studies [37].

X-ray diffraction patterns recroded form BGO precursor material showed somewhat smaller d(001) value compared to HGO and TGO (Fig. S3) in a good agreement with earlier published data [37,38,49]. Smaller averaged interlayer distance provided by d(001) is related to less defect nature of BGO.

### 3.2 Sorption of Am(III), U(VI) and Cs(I) onto different GOs

The pH dependence of sorption is a powerful source of information to understand the mechanism of reaction. The sorption data for Am(III) and U(VI) are presented in Figure 1. Sorption of U(VI) and Am(III) is dependent on pH values which is an indirect indication of inner-sphere complex formation [52]. At the same time, sorption of Am(III) takes place at lower pH values compare to U(VI). The charge of the Am(III) cation is +3, while U(VI) present in solution is in the form of uranyl cation $UO_2^{2+}$ with a total charge of +2 and effective charge of +3.2 [53]. It was previously shown that during sorption onto iron oxides such as hematite and goethite, U(VI) sorbed at lower pH values compare to Am(III) [54]. While in case of GO and other carbon nanomaterials, like nanodiamonds, [55] Am(III) sorbed more efficiently (at lower pH) than U(VI). During sorption on iron oxide interaction with hydroxyl groups take place. Similarly, the sorption results for the first hydrolysis constants for Am(III) and U(VI) are -5.25 and -7.2, respectively. The differences between the stability constants of Am(III) and U(VI) carboxylates is smaller and in some cases, for example with EDTA the formation of complexes with Am(III) is more favorable than for U(VI) [56]. Therefore, the higher adsorption of Am(III) than U(VI) on GO is evidence that the interaction of actinides with GO is more like the interaction with organic substances than with hydroxyl groups of inorganic minerals. The sorption for studied GOs is different, with BGO showing lower sorption of U(VI) and Am(III).

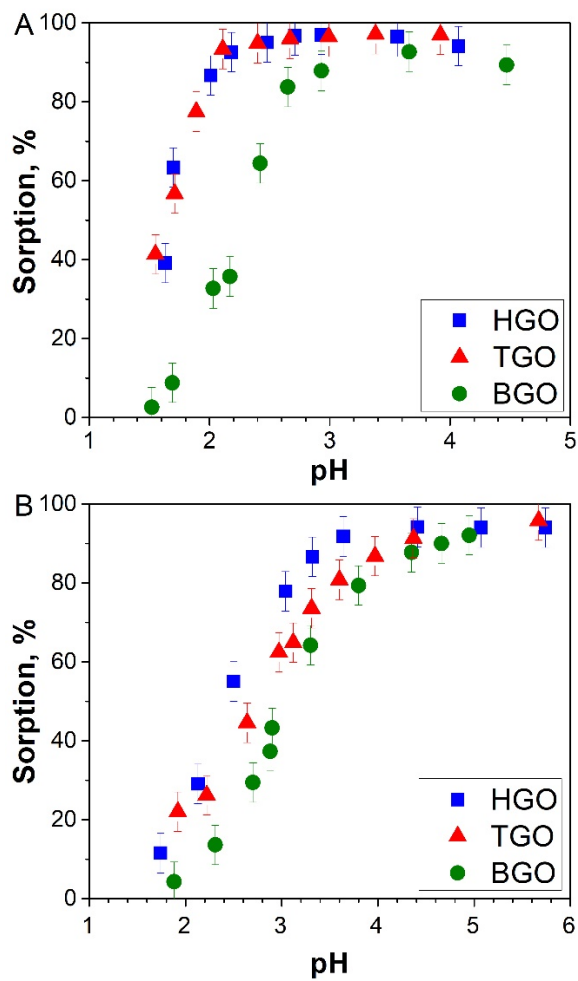

**Figure 1.** pH sorption edges for Am(III) (A) and U(VI) (B) in case of different GOs. ([GO] = 0.07g/L, [Am(III)]=$5.7 \cdot 10^{-10}$ M, [U(VI)]=$1.2 \cdot 10^{-7}$ M, I = 0.01 M).

The sorption capacities of GOs were compared using experiments in solutions with high concentrations of competing cations (saturation mode). $Al^{3+}$, $Ca^{2+}$, and $Na^+$ were chosen as competing cations for Am(III), U(VI), and Cs(I), respectively. The concentration of competing cations was high enough to saturate the GO surface. BGO had the lowest sorption values for all three radionuclides while the sorption on TGO and HGO is similar (Fig. 2). It can be concluded that TGO and HGO have higher affinity towards radionuclides both at trace level concentrations and at concentrations at which the surface is close to saturation.

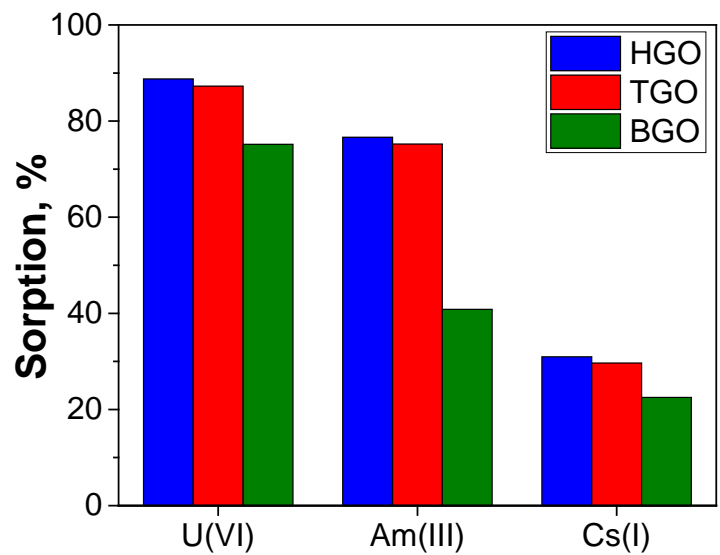

**Figure 2.** Sorption of radionuclides on different GO samples in the presence of competing cations ([GO] = 0.7g/L, [Al$^{3+}$] = [Ca$^{2+}$] = 1.3 g/L, [Na$^+$] = 0.2 g/L, pH(Am(III), U(VI)) = 3.5, pH(Cs(I)) = 5.6).

### 3.3 Characterization of GOs after sorption.

Different spectroscopic and microscopic techniques were used to reveal the mechanism of interaction of radionuclides with the studied GOs, including XPS, ATR-FTIR, EXAFS, HERFD-XANES and HRTEM. The analyzed samples are listed in Table S1.

The XPS lines from the sorbed metal cations are clearly seen in the spectra of GOs after radionuclide sorption and vacuum drying (Figure 3). The speciation of Eu(III) and U(VI) in TGO and HGO is similar while in the case of BGO, a small shift of ~0.5 eV was observed in the position of the lines. This shift indicates that the chemical bonding of cations with the functional group of BGO is slightly different than that in HGO and TGO. Spectra of Cs(I) onto all three GOs is similar. Cs(I) sorbed onto GOs with formation of outer-sphere complexes. This interaction is much weaker, which results in lower confidence due to the small XPS changes. At the same time the influence of the sorption process is less obvious in the carbon and oxygen spectra (Fig. 4, Fig. S1, Table S3).

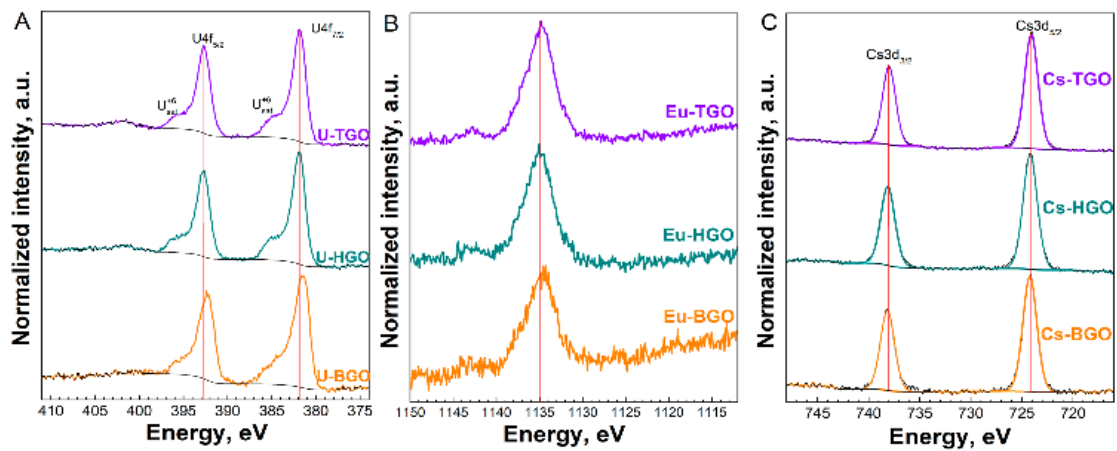

**Figure 3.** U4f (A), Eu3d$_{5/2}$ (B), and Cs3d (C) XPS spectra of cations sorbed by TGO, HGO and BGO.

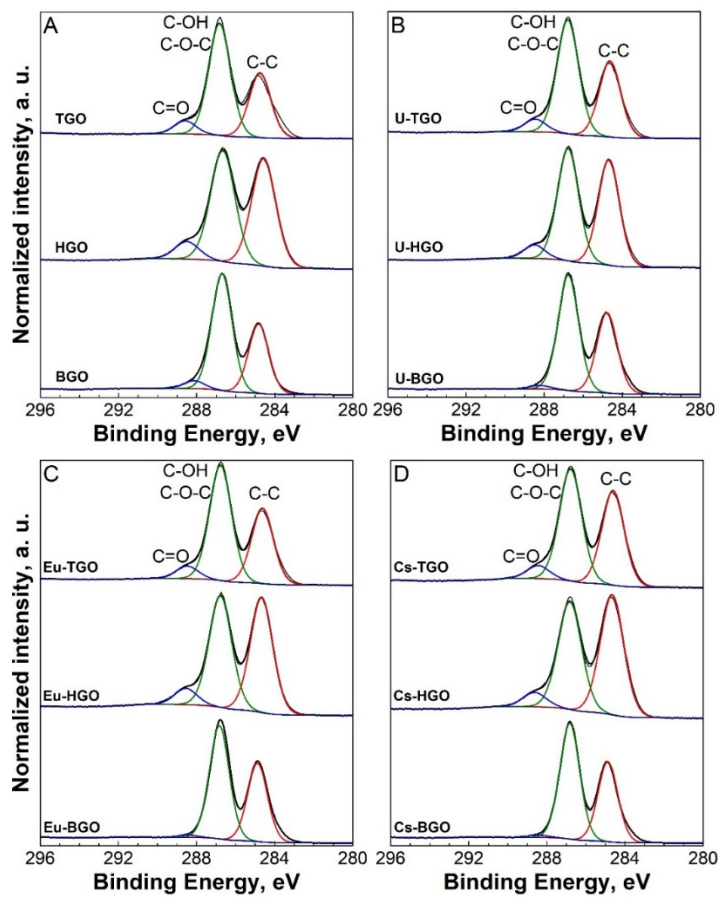

**Figure 4.** C1s XPS spectra of TGO, HGO and BGO before (A) and after sorption experiments with: (B) – U(VI), (C) – Eu(III), (D) – Cs(I). Assignment of XPS peak is according to ref.[57]

ATR-FTIR spectra of GO significantly changed as a result of Eu(III) and U(VI) sorption (Fig. 5). Less significant changes were observed in FTIR spectra as a result of Cs(I) sorption for all studied GO materials since sorption of Cs(I) on GO is relatively weak. Previously it was shown that the mechanism of interaction of Cs(I) with GO is ion exchange [24]. There is no shift in vibration modes of C-O bonds in the ATR-FTIR spectra of Cs-GO. This can be considered as a confirmation of the weak bonding of Cs(I) to GO and formation of outer-sphere complexes.

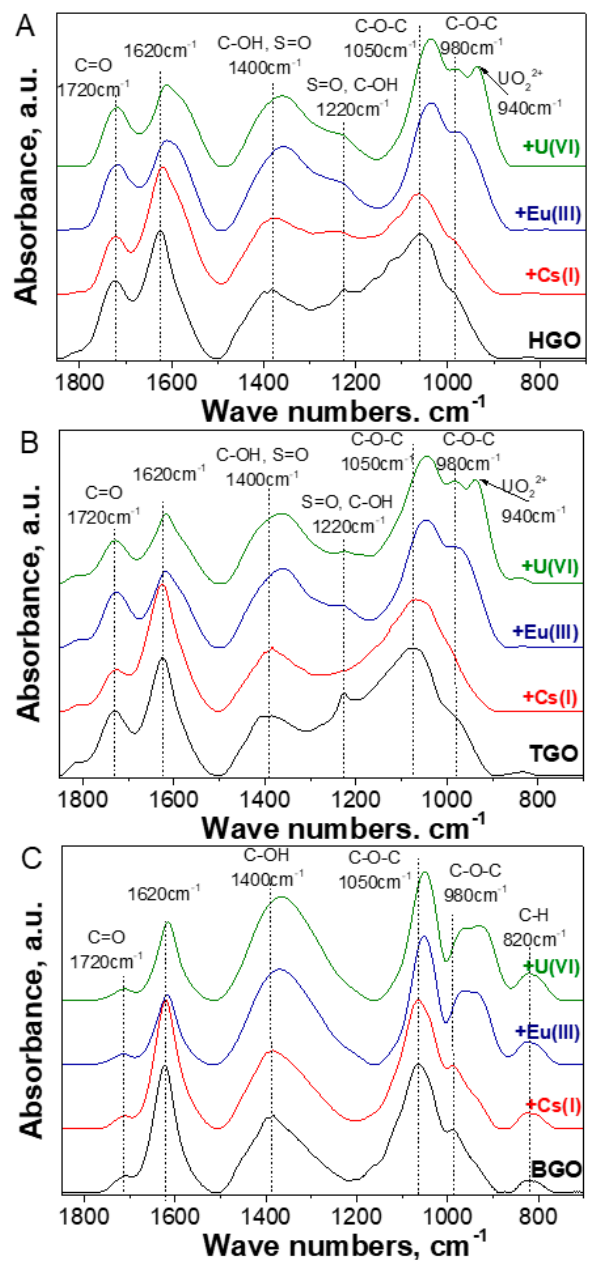

**Figure 5.** ATR-FTIR spectra of (A) – HGO, (B) – TGO and (C) – BGO before and after U(VI), Eu(III) and Cs(I) sorption.

The stretching mode of the sulfate group is found in the ATR-FTIR spectra of HGO and TGO at 1225 cm$^{-1}$[58]. This peak becomes much weaker after cation sorption. Also the XPS data show that sulfur content in the HGO and TGO samples decreases after sorption (Table S4). Therefore, the changes in the sulfur contribution to the ATR-FTIR spectra may be explained by removal of sulfate groups from the GO. The presence of S-containing groups in the GO structure or as an impurities was discussed previously [46,47,58,59]. It was shown that sulfate groups can be covalently attached to GO sheets synthesized by Hummers's method (or its modifications). Different data are available in literature regarding the stability of this species, but some examples of partial removal of sulfate from GO were shown [59,60].

The results of the present work indicate that ~70% of sulfur was removed during the sorption. Therefore, it is unlikely that most of the actinides are sorbed by sulfate groups present in the GO. The sorption mechanism related to interaction with sulfur groups was proposed previously in several works [33,61–63].

ATR-FTIR spectra of BGO with sorbed Eu(III) and U(VI) show a shift of the hydroxyl (C-OH 1400 cm$^{-1}$) and epoxy (C-O-C 1050, 980 cm$^{-1}$) vibrations by ~ 10 cm$^{-1}$. The shift of these vibrations are much higher (~30 cm$^{-1}$) in the case of Eu(III) and U(VI) sorption on TGO and HGO. Lower sorption of radionuclides by BGO correlates with relatively smaller changes in FTIR spectra after sorption.

As a result of the interaction of TGO functional groups with Eu(III) and U(VI), the peak at ~1620 cm$^{-1}$ becomes asymmetric with a clear shoulder appearing at 1580 cm$^{-1}$. This shoulder is likely to originate from carboxylates [64]. Remarkably, this shoulder is almost absent in spectra of BGO (Figure 5). That is in agreement with the less defected state of BGO and lower number of carboxylic groups in this material. Note that interpretation of 1620 cm$^{-1}$ band is controversial at the moment. Some authors attribute this peak to C=C vibrations, others to water bending. Some strong arguments supporting water as a main source of this peak were presented [20,45].

For the interpretation of the results of EXAFS and HERFD-XANES, the DFT geometry optimization of U(VI) complexes with different possible functional groups on GO was done. The interaction of U(VI) with different edge groups are presented in Fig. 6 A-E. The possible U(VI) binding above the surface of GO is presented in Fig. 6 F.

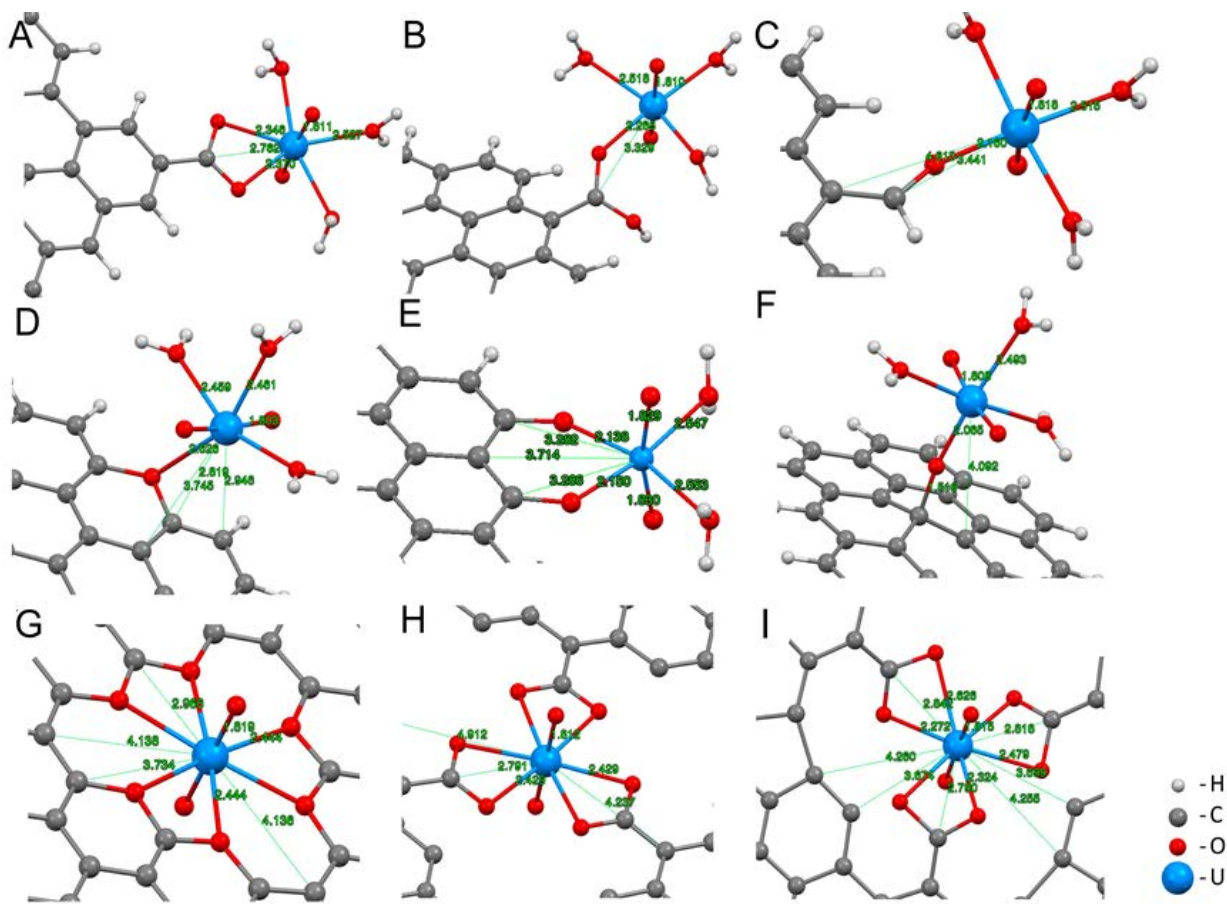

**Figure 6**. Results of DFT optimization U(VI) complexes with different functional groups on the GO.

Formation of hole defects in the GO structure is also possible especially during over-oxidation or thermal treatment. Different functional groups fill the edges of the hole forming various options for interacting with radionuclides. The examples of such interaction are presented in Fig. 6 G-I. The fitting of experimental EXAFS spectra was made on the basis of the received DFT geometry optimization. DFT geometry optimization of Eu(III) binding is presented in Fig. S4.

Experimental EXAFS spectra of U(VI) and Eu(III) sorbed by GO and its fittings are presented in Fig. 7. Structural parameters obtained from EXAFS data fitting are shown in Table 1. In U(VI) spectra, the first peak of the Fourier transform at 1.3-1.4 Å is the U-O shell and can be fitted with two oxygen atoms, indicating that the U(VI) is adsorbed on the GO surface as the uranyl ($UO_2^{2+}$) ion. The next peak corresponds to equatorial oxygen atoms and is fitted by two subshells with resulting coordination numbers of 2.3-2.5 and 2.6-4.2. The first subshell is ~2.21-2.22 Å.

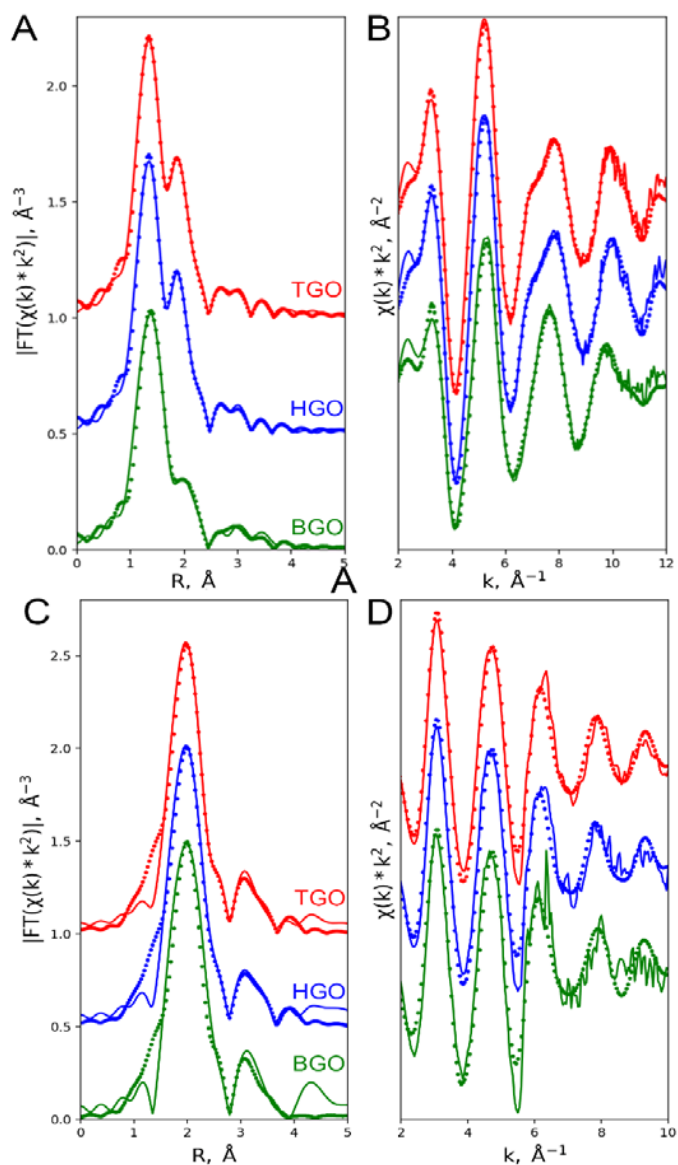

**Figure 7.** EXAFS spectra for (A,B) – U(VI) and (C,D) – Eu(III) sorbed onto different GOs. (A,C) – absolute values of Fourier transformations of the spectra, (B,D) – oscillation part of the EXAFS spectra. Dots represents the experimental data; lines are the fitting of the spectra.

Following the DFT calculation, this distance can be assigned to chemically bonded oxygen atoms from the oxygen-containing functional groups of GO. The second subshell at 2.39-2.40 Å corresponds to oxygen atoms from water molecules. A shoulder at ~2.3 Å in the FT of EXAFS spectra could be fitted by carbon atoms at ~2.9 Å and ~2.7 Å in the case of TGO and HGO samples. These carbon atoms may derive from the bidentantate U bound with functional groups (Fig. 6A). According to our DFT calculations, there are two U – C distances related to COO- group: ~2.7 Å (Fig. 6A) and ~2.9 Å (Fig. 6 G-I) located on the edge and in the holes of the GO structure, respectively.

The rest of the peaks can be attributed to C atoms which correspond to GO sheet. In the range of 3-4 Å, the EXAFS spectra of the three GO samples are characterized by carbon coordination shells at ≈3.77 Å and ≈3.96 Å and coordination numbers which are very close to each other. According to the DFT calculations, such carbon coordination shells could be assigned to carbon atoms of the planar GO sheet. It should be noted that radii and coordination numbers of distant carbon shells are very close to each other for the studied samples. Inspection of distant coordination shells for U+BGO, U+TGO and U+HGO suggest that the position of U atom in the graphene sheet are sample independant.

**Table 1** Structural parameters around uranium and europium onto GO samples derived from EXAFS analyses. For the GO samples, k ranges from 3 to 12 Å$^{-1}$ and FT from 1.2

| Sample | | CN | R | σ² | ΔE | R-factor |
|---|---|---|---|---|---|---|
| U+TGO | U-O | 2.0 | 1.75 | 0.001 | 1.5 | 0.002 |
| | U-O | 2.5 | 2.22 | 0.002 | | |
| | U-O | 4.2 | 2.39 | 0.003 | | |
| | U-C | 0.3 | 2.71 | 0.002 | | |
| | U-C | 2.2 | 2.91 | 0.002 | | |
| | U-C | 4.2 | 3.76 | 0.003 | | |
| | U-C | 3.1 | 3.97 | 0.003 | | |
| U+HGO | U-O | 2.0 | 1.75 | 0.001 | 2.2 | 0.002 |
| | U-O | 2.3 | 2.22 | 0.002 | | |
| | U-O | 4.1 | 2.39 | 0.003 | | |
| | U-C | 0.6 | 2.68 | 0.002 | | |
| | U-C | 2.6 | 2.91 | 0.002 | | |
| | U-C | 4.1 | 3.76 | 0.003 | | |
| | U-C | 3.7 | 3.95 | 0.003 | | |
| U+BGO | U-O | 2.0 | 1.76 | 0.001 | 2.8 | 0.006 |
| | U-O | 2.3 | 2.21 | 0.002 | | |
| | U-O | 2.6 | 2.40 | 0.004 | | |
| | U-C | 1.1 | 2.93 | 0.002 | | |
| | U-C | 3.0 | 3.77 | 0.003 | | |
| | U-C | 3.4 | 3.97 | 0.003 | | |
| Eu+TGO | Eu-O | 5.6 | 2.41 | 0.003 | 12.8 | 0.02 |
| | Eu-O | 4.8 | 2.54 | 0.003 | | |
| | Eu-C | 4.9 | 3.16 | 0.009 | | |
| | Eu-C | 4.3 | 3.60 | 0.008 | | |
| | Eu-C | 7.6 | 4.15 | 0.008 | | |
| | Eu-C | 5.8 | 4.49 | 0.008 | | |
| Eu+HGO | Eu-O | 5.6 | 2.41 | 0.003 | 12.9 | 0.03 |
| | Eu-O | 5.0 | 2.55 | 0.003 | | |
| | Eu-C | 4.6 | 3.16 | 0.007 | | |
| | Eu-C | 4.0 | 3.63 | 0.007 | | |
| | Eu-C | 7.8 | 4.16 | 0.007 | | |
| | Eu-C | 5.9 | 4.50 | 0.007 | | |
| Eu+BGO | Eu-O | 5.8 | 2.42 | 0.003 | 12.7 | 0.04 |
| | Eu-O | 4.2 | 2.55 | 0.003 | | |
| | Eu-C | 6.0 | 3.17 | 0.008 | | |
| | Eu-C | 6.6 | 3.61 | 0.008 | | |
| | Eu-C | 5.4 | 4.14 | 0.008 | | |

It was proposed earlier that carboxyl groups can be located only on the edges of the GO sheets. However recent papers confirm that carboxyl groups can be located on the edges of nanometer scale holes or vacancy points in the GO sheets [37,65]. The present EXAFS data prove that U(VI) preferentially interacts with functional groups located in the holes of GO sheet. Moreover, in the case of TGO and HGO, the U(VI) predominantly interacts with carboxyl groups. Species of U(VI) sorbed onto BGO are somewhat different. The U-C distance at 2.7 Å is not observed and the coordination number of the C atoms at 2.9 Å is significantly decreases compared to U-HGO and U-TGO, indicating that the interaction of U(VI) is mainly with hydroxyl groups.

EXAFS spectra and their fitting for Eu(III) sorbed onto HGO, TGO and BGO are quite similar. The interaction of Eu(III) with carboxyl groups on all studied GO samples are in agreement with the presented EXAFS spectra. Batch sorption experiments show the significantly smaller sorption of Am(III) by BGO sample (Fig. 1,2) than with HGO and TGO. As discussed above, the BGO material contains fewer carboxyl groups compared to HGO and TGO. Therefore, the interaction of Am(III) preferentially with carboxyl groups would explain the significantly smaller sorption capacity of BGO.

HERFD-XANES spectra collected at U L3 edge for U(VI) sorbed on TGO, HGO and BGO are show in Figure 8. The spectra for U(VI) in TGO, HGO and BGO are identical, indicating that U(VI) has the same local coordination in all samples. To facilitate the observation of the spectral differences, one of the samples, U+HGO, is compared with the uranyl nitrate ($UO_2(NO_3)_2 \times 6H_2O$) reference shown in the top of Fig. 8. The absorption edge in sorbed U(VI) is shifted towards low incident energy of ~1 eV compared to the absorption edge of uranyl reference. The post-edge region shows also sizeable differences in spectra for the U-GO samples and the $UO_2(NO_3)_2$ reference in correspondence of feature A and B. These spectral differences reflect the different local environments of U(VI) sorbed on GO compared to uranyl and theoretical simulations can be used to check which structures are compatible with experimental data. The DFT optimized structures of U(VI) sorbed at the edge and in holes of the GO sheet were used as input for FDMNES calculations and the results were compared to the simulation of uranyl ion. All structures with U(VI) at the edge of the GO sheet gave similar results. Therefore, only spectra corresponding to the structure in Fig. 6A, E are shown. Although the general trend in the post-edge is reproduced by this specific U(VI) configuration in GO, the intensity of feature B is underestimated compared to the experimental data. For structures with U(VI) occupying a hole of the GO sheet, the variations observed in the post-edge region are very well reproduced, both in terms of intensity and energy position, therefore this local coordination gives better agreement with experimental data. The differences between structures with U(VI) occupying a hole appear the pre-edge and main edge region, in particular the structure shown in Fig. 6H has a more intense pre-edge peak and has the main absorption edge slightly shifted to lower energy. We stress that the absorption pre-edge and edge regions are the most difficult to model and the simulation results in these regions should only be taken as suggestive of possible trends. The pre-edge of the uranyl ion is indeed very nicely resolved in HERFD-XANES when acquired with higher resolution that the one of the present data [66]. Acquiring data with higher resolution could unveil differences in the absorption edge region that could help to distinguish between different local coordinations of U(VI). In general, simulations of HERFD-XANES based on the DFT optimized structures confirmed that U(VI) is not sorbed at the edge of the GO sheet, but it is rather occupying a hole.

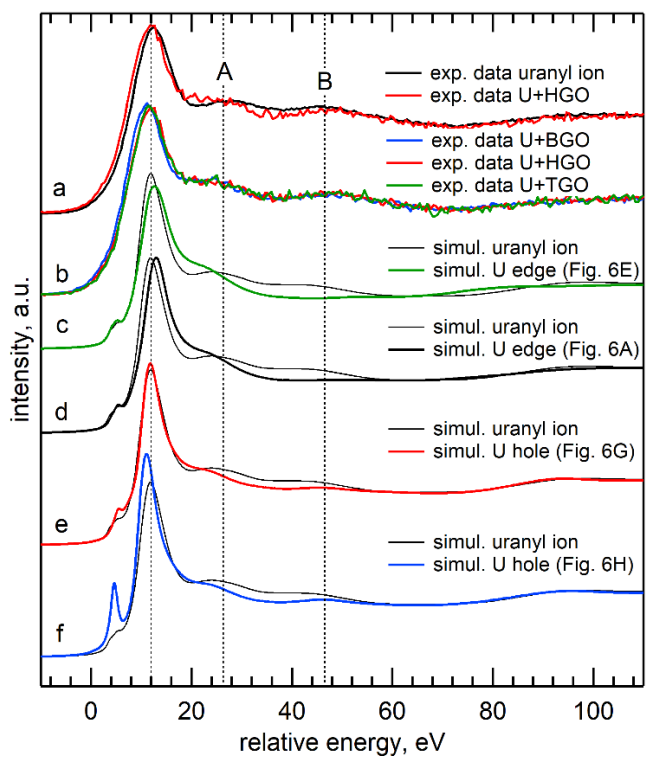

**Figure 8.** From top to bottom: a) U L$_3$ edge HERFD-XANES data of UO$_2$(NO$_3$)$_2$×6H$_2$O reference compared to U(VI) sorbed onto HGO; b) U L$_3$ edge HERFD-XANES of U in BGO, HGO and TGO. FDMNES simulations of uranyl reference (black thin line) compared with simulation on three DFT optimized structures: in c) the one shown in Fig. 6E and d) in Fig. 6A with U at the edge of a flake, in d) and e) the structure shown in Fig. 6G and 6H with U in a hole of GO.

High resolution STEM was used to study the distribution of sorbed cations on the GO sheets. The images collected using this method provide information about the distribution of Eu(III) on the BGO, HGO and TGO surfaces (Fig. 9). The white colour on Fig.9A,C,D,E corresponds to high atomic weight Eu atoms. In the case of BGO, Eu(III) is predominantly located on the wrinkles of the GO. In case of TGO and HGO, Eu is well-dispersed, with no preferred sorption on the edge sites of the GO. This is in agreement with the suggestions that Eu(III) and U(VI) interact with carboxyl groups located in holes and defects on the GO sheet. Homogeneous distribution of metal atoms over the surface of HGO and TGO flakes suggests that vacancy defects are an inherent property of GO produced by the permanganate route, in agreement with earlier studies[67].

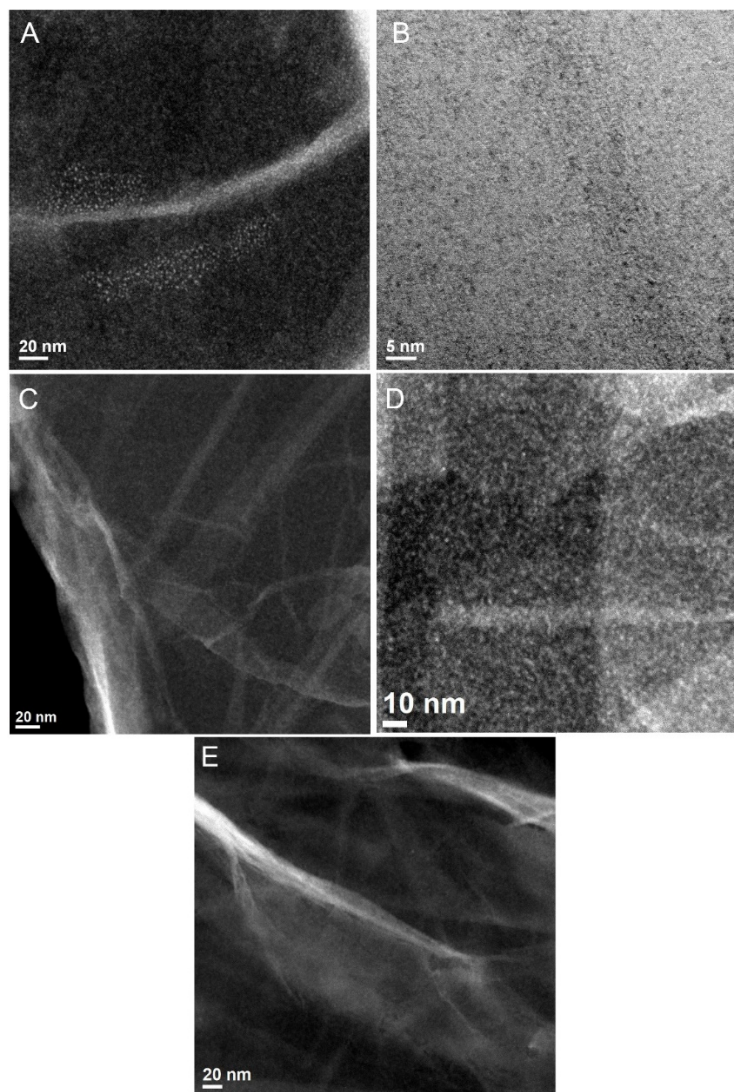

**Figure 9.** TEM images of Eu(III) onto (A,B) TGO, (C,D) HGO and (F) BGO in (A,C,D,E) STEM and (B,D) HRTEM modes.

## 4. CONCLUSIONS

Sorption properties of GO synthesized by Hummers's, Brodie's and Tour's methods towards different radionuclides was compared. It was observed that BGO has lower sorption capacity and lower activity compared to HGO and TGO in the case of Am(III)/Eu(III) and U(VI). The lower sorption of the radionuclides by BGO correlates with the smaller relative amounts of carboxyl groups and smaller overall oxidation degree of this material. Analysis of data obtained using various spectroscopic techniques suggests that Am(III)/Eu(III) and U(VI) mostly interact with carboxylic groups. Analysis of EXAFS and HERFD-XANES spectra combined with DFT simulations showed that radionuclides predominantly occupy vacancy defects in GO sheets. This conclusion is also confirmed by direct high resolution imaging which reveals homogeneous distribution of metal atoms over the HGO and TGO surface rather than along the edges of GO flakes. The lower sorption of radionuclides by BGO is then explained by less defected nature of this material compared to GO synthesized by the permanganate route.

Improvements of the radionuclides sorption capacity can possibly be achieved using GO materials with an intentionally high number of defects. Therefore, our study provides new and highly promising strategy for designing materials with improved sorption capacity towards radionuclides, e.g. by using perforated GO.

### ACKNOWLEDGEMENTS

Work was supported by the Russian Science Foundation (project 14-13-01279). Experimental studies were partially performed on the equipment acquired with the funding of M.V. Lomonosov Moscow State University Program of Development. We also acknowledge support from HZDR for beamtime allocation. A.T. acknowledge funding from the European Union Horizon 2020 research and innovation program under grant agreements No. 696656 and No. 785219. L.A and K.K acknowledge funding from European Research Council (ERC) under the ERC Starting Grant agreement N 759696 (acronym TOP). The work at Rice University was supported by the Air Force Office of Scientific Research (FA9550-14-1-0111). DFT geometry optimization has been carried out using computing resources of the federal collective usage center Complex for Simulation and Data Processing for Mega-science Facilities at NRC "Kurchatov Institute", http://ckp.nrcki.ru/.

**Supporting Information**. XPS, XRD, FTIR data and results of DFT optimization for Eu(III)-GO interaction.

competitive binding of several metal cations by graphene oxide reveals the quantity and spatial distribution of carboxyl groups on its surface, Phys. Chem. Chem. Phys. 20 (2018) 2320–2329. doi:10.1039/c7cp07055a.